\documentclass[a4paper,twocolumn,british,prl,superscriptaddress,longbibliography,floatfix,notitlepage]{revtex4-1}
\usepackage[T1]{fontenc}
\usepackage[latin9]{inputenc}
\usepackage{amsmath}
\usepackage{amssymb}
\usepackage{graphicx}

\makeatletter

\pdfpageheight\paperheight
\pdfpagewidth\paperwidth

\usepackage{xcolor}
\definecolor{darkblue}{rgb}{0.1,0.2,0.6} 
\definecolor{lightblue}{rgb}{0.1,0.1,1.0}
\definecolor{darkred}{rgb}{0.8,0.1,0.2}
\usepackage[colorlinks,citecolor=lightblue,linkcolor=darkblue,urlcolor=lightblue] {hyperref}
\usepackage[british]{babel}

\makeatother
\begin{document}
\global\long\def\bra#1{\left\langle #1\right|}%
\global\long\def\ket#1{\left|#1\right\rangle }%

\title{Locality versus Fock-space structure in East-type models}
\author{Achilleas Lazarides}
\affiliation{Loughborough University, Loughborough, Leicestershire LE11 3TU, UK}
\begin{abstract}
   Local kinetic constraints in quantum many-body systems can generate slow dynamics or complete many-body localisation. Here we focus on a modification of the quantum East model: Inspired by random matrix theory, we randomise the connectivity in Fock space (rendering it nonlocal in real space) while preserving its organisation into neighbouring magnetisation sectors. We find that there is still a transition between two distinct phases, one delocalised and the other localised. We conclude that, for East-type constrained models, the essential ingredient is the structure of the graph in Fock space rather than geometric locality of spin flips. 
\end{abstract}
\maketitle

\section{Introduction}

\begin{figure}[t]
  \centering
  \includegraphics[width=\columnwidth]{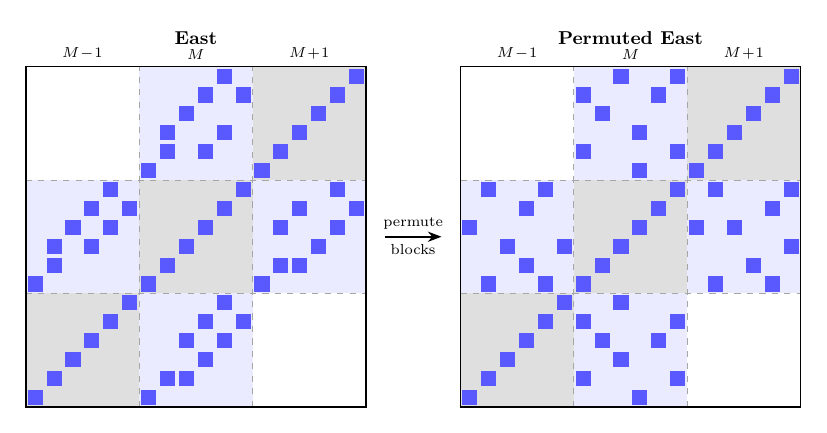}
  \caption{Schematic of the permuted East construction. Each matrix represents the Hamiltonian in the Fock-state basis, with rows and columns ordered so that states with the same number of up spins $M$ form contiguous blocks (shaded grey on the diagonal). The diagonal entries (blue dots on the main diagonal) represent the on-site disorder, which is identical in both models. Only nearest-neighbour blocks ($M \leftrightarrow M\pm 1$, light blue) carry off-diagonal entries, reflecting the single-spin-flip structure. In the East model (left) the nonzero off-diagonal entries within each block follow the structured, constraint-driven pattern imposed by real-space locality. In the permuted East model (right) the entries in each off-diagonal block are independently reshuffled: the number of connections between each pair of sectors is preserved, but their arrangement is randomised, removing real-space locality while retaining the block structure. See Fig.~\ref{fig:composite} for the full matrix patterns and corresponding Fock-space graphs.}
  \label{fig:schematic}
\end{figure}

Ergodicity is a fundamental concept underpinning the applicability of  statistical mechanics. Almost everything is ergodic, so exceptions are interesting. More practically, the absence of ergodicity is a key ingredient in the construction of quantum simulators and quantum computers: Many-body quantum systems that perform work or computation are often non-adiabatically driven and, in the ergodic case, such systems will thermalise to a featureless infinite-temperature state~\cite{Lazarides:2014ie}. Another example is that of novel phases of matter, such as time crystals, which also require the absence of ergodicity~\cite{Khemani2016}. Thus understanding the dynamics, and thermalisation, of many-body systems is of both fundamental and practical importance. 

\begin{figure*}[t]
  \centering
  \includegraphics[width=2.0\columnwidth]{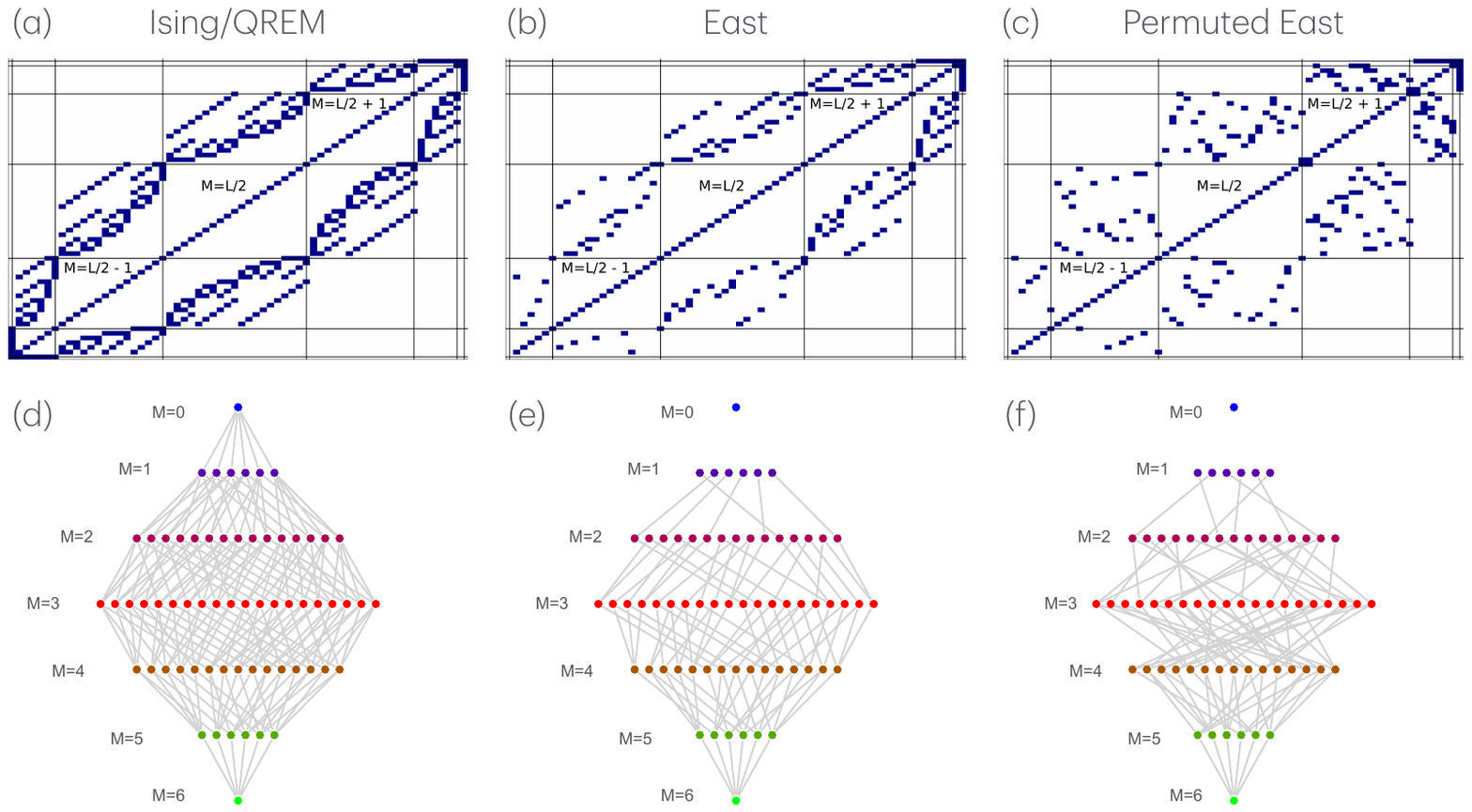}
  \caption{Panels (a-c) show the locations of the nonzero matrix elements in the Hamiltonians of the standard Ising, quantum East and permuted East models. Panels (d-f) show the graphs described by the corresponding adjacency matrices, obtained from the Hamiltonians in panels (a-c) via Eq.~\eqref{eq:adjacency-from-h}. The horizontal and vertical axes in panels (a-c) correspond to Fock state labels, ordered so that Fock states with the same number of up spins $M=\sum_i n_i$ are next to each other. Each diagonal block corresponds to a set of Fock states with the same number of up spins $M$, as indicated for three blocks in each panel, and is connected only to the neighbouring blocks $M\pm 1$, reflecting the fact that the Ising model only includes single spin flips. Each site in the East model is connected to a subset of the sites to which the same site is connected in the Ising model, because the effect of the constraint in Eq.~\ref{eq:eastrem} is to forbid some spin flips. By contrast, in the permuted East model the total number of connections between each neighbouring pair of magnetisation sectors is unchanged, but the connectivity of individual Fock states varies. The ladder-like organisation into magnetisation sectors is therefore preserved, while the detailed connectivity within each pair of neighbouring sectors is scrambled.}
  \label{fig:composite}
\end{figure*}

A prominent and well-studied case of broken ergodicity is that of many-body localisation (MBL)~\cite{Oganesyan:2007ex,Nandkishore2014,Gornyi2005,Basko2006}. More recently, however, significant effort has been devoted to finding non-disordered systems breaking ergodicity~\cite{Carleo2012,DeRoeck2014Scenario,Grover2014,Schiulaz2015,Papic2015,Barbiero2015,Yao2016,Smith2017,Mondaini2017,Yarloo2018,Schulz2019,vanHorssen:2015ts,Hickey2016,Shiraishi2017,Lan2018,Feldmeier2019,vanNieuwenburg2019,Zisling2022,Doggen2021,Kloss2023}.

Some constrained models display Hilbert space fragmentation (HSF), for example, certain fracton or dipole-conserving models~\cite{Prem2017,Nandkishore2019Fractons,Pretko2020FractonPhases,Chamon2005,Haah2011,Yoshida2013,Vijay2015,Vijay2016,Pretko2018Elasticity,Pretko2017Subdimensional,Pretko2018GaugePrinciple,Pretko2017Witten,Williamson2019,Sala2020,Khemani2020Shattering,Rakovszky2020StatisticalLocalization,Scherg2021, Feldmeier2020,Morningstar2020}. Others break ergodicity only weakly (in their largest connected sector of Hilbert space) but in novel ways, such as the PXP model~\cite{Turner2018,papic_weak_2021,Serbyn2018,Ho2019,Lin2019}. Here, we are instead interested in models with a connected Hilbert space that localise, thus break ergodicity, through mechanisms other than real-space disorder. Most relevant to this work is the quantum East model~\cite{vanHorssen:2015ts,Pancotti2019,menzler_graph_2025,das_ground_2026}. This is a model without HSF which shows a localisation-delocalisation transition in the ground state, and a related fast-to-slow transition in the bulk of its spectrum.

In general, studies of ergodicity and its breaking are technically difficult, and even if a problem is solvable it is unclear which features of the solution are generic and which specific to that system. One of the first ways around this was the introduction by Wigner of random matrix theory, in which random matrices, rather than actual Hamiltonians, are studied~\cite{Wigner1955,Wigner1957,Wigner1958}. This approach has been very successful. The general classes of orthogonal, unitary and similar ensembles of random matrices have been well studied in the context of many-body dynamics~\cite{DAlessio2016}. 

While these can successfully describe gross properties of chaotic systems, they fail to describe many details. For example, equilibration happens much faster than is realistic~\cite{santos_analytical_2017,torres-herrera_generic_2018}, while eigenstate properties do not depend on the energy density in the same way that they do in real systems. Another difference is that the density of states follows a semicircle law for random matrices, but generally is Gaussian for many-body systems~\cite{DAlessio2016}.

To rectify such issues, effort has been devoted to designing random matrices that mirror physical properties of Hamiltonians beyond their global symmetry. These include two-body embedded ensembles~\cite{french_validity_1970,bohigas_two-body_1971,kota_embedded_2001, vyas_random_2018} and power-law banded matrices~\cite{mirlin_transition_1996,bogomolny_eigenfunction_2011,buijsman_power-law_2025}, which embody locality properties of Hamiltonians while otherwise randomising them.

A few years ago, aiming to find a model with localisation in the bulk of the spectrum, we studied an East-type model and found that ergodicity is indeed broken there due to the constraints~\cite{Roy2019a}. This model has disorder, but uncorrelated between the different Fock states. This type of disorder does not, in general, cause localisation~\cite{roy_fock-space_2020}. Indeed, in the so-called Quantum Random Energy Model (QREM), which has such disorder plus unconstrained spin flips $\sum_j\sigma_j^x$~\cite{Laumann:2014ju}, all states except those at the spectral edges are delocalised. In Ref.~\cite{Roy2019a} we added East-type constraints to the QREM to obtain the EastREM, so that a spin can only flip if its right-neighbour is pointing up, and found that there is a critical hopping rate below which \emph{all} states are localised. In other words, the rate at which a spin flips depends upon the state of its neighbour.

Here, drawing inspiration from random matrix theory, we write the Hamiltonian in matrix form in the product-state basis, then progressively remove structure from this matrix and see whether the defining properties of the East model are retained. In particular we recognise that the Hamiltonian of the East model in the Fock state basis consists of blocks corresponding to states with $M$ spins pointing up, each connected to the neighbouring blocks with $M\pm 1$ spins pointing up. Exploiting this, we construct another model, the \emph{permuted} East model, in which the ladder-like organisation into magnetisation sectors is retained while the connections between neighbouring blocks are randomised. Thus real-space structure is scrambled while the coarse Fock-space organisation is preserved. Using the East model as a previously-studied benchmark, we show that there is the same transition in the permuted model as in the original one. The construction is illustrated schematically in Fig.~\ref{fig:schematic}; see Fig.~\ref{fig:composite} for the full matrix patterns and Fock-space graphs.

At a high level, our results show that scrambling the real-space structure of the East model while preserving its organisation into neighbouring magnetisation sectors leaves the qualitative phenomenology intact. Using dynamical participation ratios and eigenstate Shannon entropy, we find that the permuted East model displays the same transition as the original East model: for small $s$ the system behaves ergodically and states spread over a large fraction of Fock space, while for larger $s$ this spreading is suppressed and the eigenstates become increasingly non-ergodic. Finite-size analyses of the Shannon-entropy fluctuations indicate that in both models this change is consistent with a transition at a finite value of $s$. These results suggest that for East-type constrained systems the key ingredient behind slow dynamics and localisation is not spatial locality of the spin flips, but the organisation of connectivity in Fock space.

A closely related perspective was developed recently in Ref.~\cite{menzler_graph_2025}, where the usual quantum East model is also analysed from a graph-theoretic viewpoint. Both works focus on perturbing the structure of the model in Fock space, but do so in different ways. In Ref.~\cite{menzler_graph_2025}, the connectivity of the quantum East graph is kept fixed, while the diagonal terms are modified, corresponding to detuning the energies of different Fock states. Here we instead keep the broad ladder-like organisation in magnetisation sectors intact, randomising the off-diagonal connectivity between neighbouring sectors. The two works are therefore complementary: taken together, they indicate that both the hierarchical graph structure and the way it is energetically weighted play an important role in the anomalous dynamics of East-type systems.

In what follows, we first define the permuted East model and clarify which structural properties of the East Hamiltonian it preserves. We then compare the two models via diagnostics of Fock-space spreading and eigenstate structure, and find that they display very similar phenomenology.

\section{Model Hamiltonians, adjacency matrices and graphs}

\begin{figure*}[t]
  \centering
  \includegraphics[width=1\textwidth]{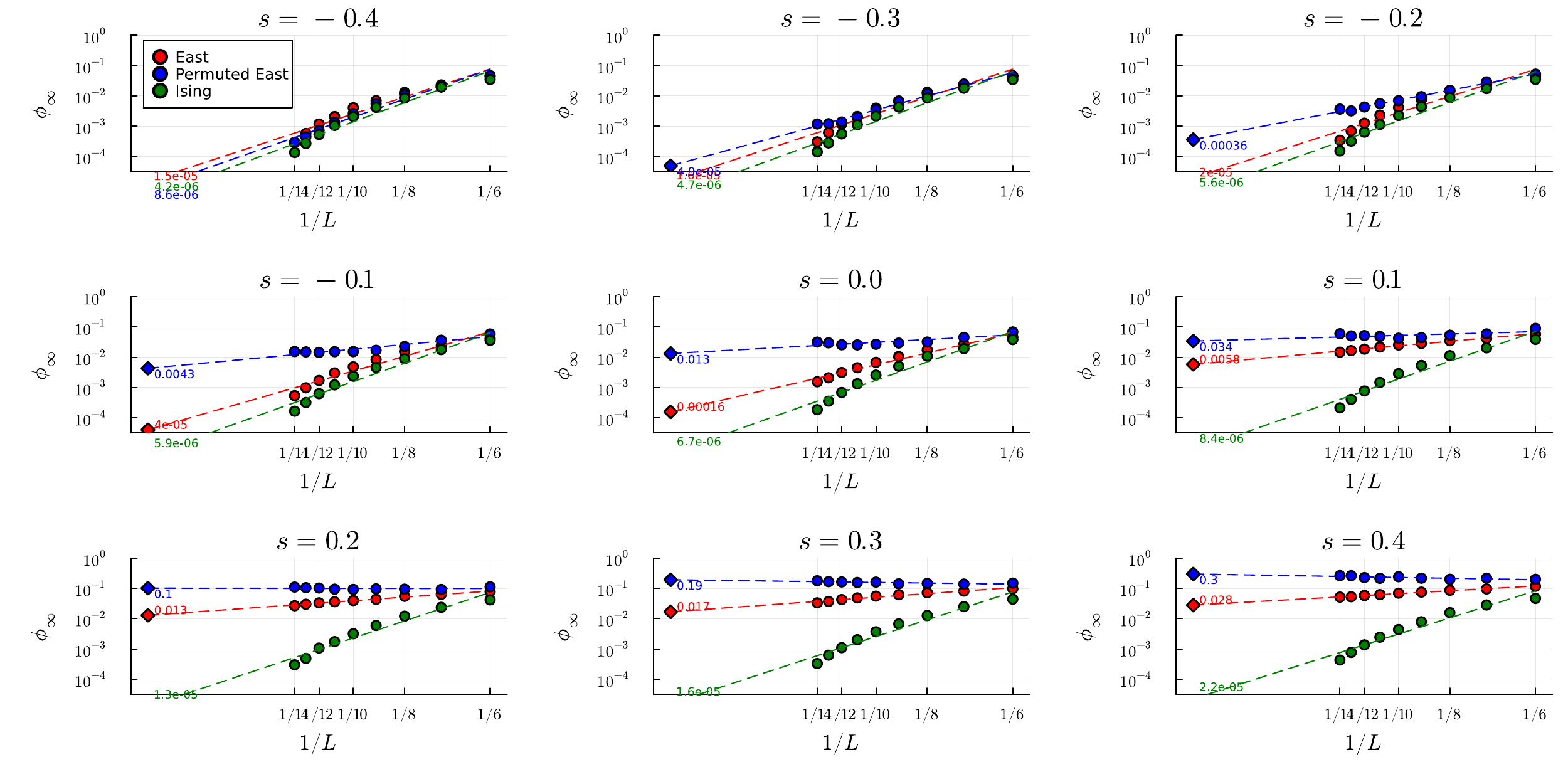}
  \caption{Infinite-time average of the dynamical participation ratio $\phi_\infty = \lim_{t\to\infty}\phi(t)$, defined in Eq.~\eqref{eq:phi-defn}, for all three models with $w = 0$ (no Fock-space disorder) and $s$ ranging from $-0.4$ to $0.4$ in steps of $0.1$. Initial states are the $L$ single-spin-up product states; $\phi_\infty$ is averaged over these initial states and, for the permuted East model, also over realisations of the random permutation. Because these are low-entanglement states in the $M=1$ magnetisation sector, they probe the low-energy part of the spectrum rather than the bulk, so the results are not representative of the full spectral average. In the fully delocalised limit $\phi_\infty = \mathcal{O}(D^{-1})$, while in the fully localised limit $\phi_\infty = \mathcal{O}(1)$. Each panel shows data for a fixed $s$ as a function of $1/L$; dashed lines are linear fits to $\log \phi_\infty$ versus $1/L$, and the diamond marker on the vertical axis indicates the extrapolated value at $1/L \to 0$ (thermodynamic limit). At $s=-0.4$ all three models extrapolate to small values consistent with delocalisation. At $s=0.4$ the two constrained models extrapolate to values consistent with localisation, while the Ising model extrapolates to a value at least three orders of magnitude smaller, consistent with remaining delocalised. The precise location of the transition or crossover between these behaviours cannot be extracted from this data due to finite-size noise.}

  \label{fig:ipr-both}
\end{figure*}

\subsection{Hamiltonians and adjacency matrices}

Any quantum spin-1/2 model can be written in terms of Fock states $\ket{n}$ as 
\begin{equation}
   H = 
   \sum_{n,m;n\neq m} T_{n,m} \ket{n}\bra{m}
   +
   \sum_m E_m \ket{m}\bra{m}.
   \label{eq:hamiltonian-general-fock}
\end{equation}
Here $T_{n,m}$ are the matrix elements for transitions between any two Fock states~\footnote{We use \emph{Fock states} to refer to computational basis states: product states of eigenstates of the $\sigma^z$ Pauli operator on each site.}, while the $E_m$ encode diagonal terms. For example, if a Hamiltonian contains a term $\sigma^x_j$ then this will result in $T_{n,m}=1$ between $\ket{n},\ket{m}$ differing by a single spin flip. 

Eq.~\eqref{eq:hamiltonian-general-fock} can be viewed as a single-particle hopping between Fock states: the basis states $\ket{n}$ play the role of sites, the off-diagonal matrix elements $T_{n,m}$ are hopping amplitudes, and the diagonal matrix elements $E_n$ are on-site energies.

For graph-theoretic purposes it is useful to separate this connectivity information from the full Hamiltonian. We therefore associate to $H$ an \emph{adjacency matrix} with entries
\begin{equation}
    A_{n,m} = (1-\delta_{n,m})\,\Theta\!\left(|H_{n,m}|\right),
    \label{eq:adjacency-from-h}
\end{equation}
where $\Theta(x)=1$ for $x>0$ and $\Theta(0)=0$. This only records which pairs of Fock states are connected by a nonzero off-diagonal matrix element, via a unit value at that entry; it does not retain the numerical values of the $T_{n,m}$ beyond that, nor the diagonal energies $E_n$. Given the adjacency matrix associated with a Hamiltonian, one can visualise the graph it describes; for the Hamiltonians shown in Fig.~\ref{fig:composite}(a-c), the corresponding graphs are shown in panels (d-f).

We will freely switch between these viewpoints (Hamiltonian, adjacency matrix, and graph).

\subsection{QREM/Ising-type model}
\label{sec:ising-type-model}

As a baseline model to orient ourselves we use the Ising-type Quantum Random Energy model (QREM), which is a model with Ising-type spin flips and a random energy for each Fock state. Its Hamiltonian is given by
\begin{equation}
   H = -\Gamma \sum_{j=1}^{L}\sigma^x_j
   +
   w\sum_n  \epsilon_n \ket{n}\bra{n}.
   \label{eq:ising}
\end{equation}
We will refer to this as the Ising or QREM model; both of these have the same connectivity structure in Fock space, with only the diagonal matrix elements differing between the two. We include this model as a baseline against which to compare the two constrained East models introduced below. For the QREM ($w\neq 0$), the phenomenology is as follows~\cite{Laumann:2014ju,Lazarides:2015jd}: The middle of the spectrum is delocalised for all values of  $\Gamma/w$, while the very edges localise at small enough $\Gamma/w$.

We label Fock states so that states with the same number of up spins $M=\sum_j n_j$ (with $n_j=\frac{1}{2}\left(1+\sigma^z_{j}\right)$) are grouped together. The nonzero matrix elements of the Ising Hamiltonian then form the block structure shown in Fig.~\ref{fig:composite}(a): diagonal blocks of dimension $\binom{L}{M}$ for each $M$, connected to adjacent blocks with $M\pm 1$ since $H$ flips a single spin. Each state connects to $L$ others, giving each block $L\binom{L}{M}$ outgoing connections -- a ragged ladder whose block dimensions scale with $L$, preventing reduction to a standard single-particle Anderson problem.

Applying Eq.~\eqref{eq:adjacency-from-h} to the Hamiltonian gives the adjacency graph shown in Fig.~\ref{fig:composite}(d). Each node is a Fock state, positioned vertically by $M$, with edges only between nodes sharing a nonzero matrix element. Edges connect only nearest-neighbour rows, each site has exactly $L$ neighbours, and the dynamics of the many-body state is equivalent to single-particle hopping on this graph.

Next we turn to two constrained versions of this model. In both we will leave the diagonal matrix elements unchanged, and modify the spin flipping term. This modifies the connectivity structure encoded in the corresponding adjacency matrices and graphs shown in Fig.~\ref{fig:composite}.

\subsection{East-type model}
\label{sec:east-type-model}

The first model combines features of the quantum East model~\cite{vanHorssen:2015ts} and the EastREM~\cite{Roy2019a} and is described by
\begin{equation}
   \begin{split}
      H =& \sum_{j=1}^{L-1} \left(
         -\mathrm{e}^{-s} \sigma^x_j n_{j+1}
         + n_j \right) 
         +
      n_L - \mathrm{e}^{-s} \sigma^x_L n_1\\
      &+
      w\sum_m  \epsilon_m \ket{m}\bra{m}.
   \end{split}
   \label{eq:eastrem}
\end{equation}
The first sum describes the quantum East model~\cite{vanHorssen:2015ts} while the second describes disorder uncorrelated among Fock sites~\cite{Roy2019a,Laumann:2014ju,Baldwin:2015ur}. This is distributed according to $p(\epsilon)=\frac{1}{\sqrt{2\pi L}} \exp\left(-\frac{\epsilon^2}{2 L}\right)$ so that its standard deviation scales with $L$ in the same way as it does in the presence of local disorder. Disorder uncorrelated between different Fock states does not result in localisation in unconstrained models~\cite{roy_fock-space_2020}.

The constraint on the spin flip disallows some flips that were allowed for the Ising model, and does not introduce new allowed spin flips. In particular, each Fock state is now connected to $M=\sum_j n_j$ others, so that the blocks become more sparsely connected as $M$ decreases. The pattern of nonzero matrix elements of the Hamiltonian is shown in Fig.~\ref{fig:composite}(b), using the same ordering of Fock states as in the Ising model (Fig.~\ref{fig:composite}(a)). There are still only matrix elements between nearest neighbour blocks, so that transitions change $M$ by $\pm 1$ by flipping a single spin, meaning that the Hamming distance between any two neighbouring states is always 1. However, each state is now connected to a subset of the states it was connected to in the Ising model, because the effect of the constraint in Eq.~\ref{eq:eastrem} is to forbid some spin flips. In terms of the graph associated with the corresponding adjacency matrix, shown in Fig.~\ref{fig:composite}(e), nodes lower down (with smaller $M$) are more sparsely connected, and the final $M=0$ node is isolated (note that apart from this node, the rest of the graph remains fully connected).

This model was extensively studied in~\cite{Roy2019a} using level spacing statistics, Fock-space participation entropies, and dynamical autocorrelations starting from product states. The central result is a localisation transition in the bulk of the spectrum at $\Gamma_c/w \approx 0.18$ (equivalently $s_c \approx 1.8$ for $w=1$), in contrast to the QREM whose bulk remains delocalised at any finite $\Gamma$~\cite{Laumann:2014ju,Baldwin:2015ur}. For $s < s_c$ the system is delocalised: eigenstates are spread over a macroscopic fraction of Fock space, level statistics follow the Wigner--Dyson distribution, and dynamical correlations decay to zero. For $s > s_c$ all eigenstates are localised, level statistics become Poisson, and dynamical autocorrelations saturate to a non-zero long-time value. In the present work we probe the same phenomenology through the dynamical participation ratio (Sec.~\ref{sec:ipr}) and Shannon entropy (Sec.~\ref{sec:entropy}), which are closely related to the diagnostics of~\cite{Roy2019a} and allow a direct comparison between the East and permuted East models.

\subsection{Permuted East model}
\label{sec:permuted-east-model}

We now introduce the central model for this work, which we will refer to as the \emph{permuted} East model. We are aiming to preserve the coarse magnetisation-sector structure, but otherwise scramble which states are connected. To this end we randomise the off-diagonal matrix elements $T_{n,m}$ of Eq.~\ref{eq:hamiltonian-general-fock} in the following way.

We first order the Fock basis by magnetisation $M$, so that states with the same value of $M$ are grouped into a common block. Let $B^{(M)}$ denote the off-diagonal block connecting the sectors with magnetisations $M$ and $M+1$. For each $M$, independently of all other blocks, we replace this block by
\[
   \tilde B^{(M)} = P_M B^{(M)} Q_M^{\mathsf T},
\]
where $P_M$ and $Q_M$ are independent random permutation matrices. The block connecting the sectors $M+1$ and $M$ is then fixed by Hermiticity, and the diagonal blocks are left unchanged. This construction preserves the magnetisation-sector decomposition, the dimensions of the blocks, the fact that couplings only occur for $\Delta M=\pm 1$, and the total number of nonzero matrix elements between each neighbouring pair of sectors. It does not preserve which individual Fock states are connected, nor the degree of each individual Fock state.


Finally, real-space locality as well as translational invariance are both lost: states connected by a matrix element still differ in magnetisation by $\pm 1$, but their Hamming distance can be larger than 1.

For a particular realisation of this model, the pattern of nonzero matrix elements of the Hamiltonian is shown in Fig.~\ref{fig:composite}(c), while the graph described by the corresponding adjacency matrix is shown in Fig.~\ref{fig:composite}(f). Both of these visually confirm that the structure remains ladder-like, with the sites organised in groups labelled by $M$ and transitions only allowed between blocks $M$ and $M \pm 1$.

\section{Numerical results}
\label{sec:numerical-results}

We compare the East and permuted East models using two diagnostics: a dynamical participation ratio, measuring spreading of the wavefunction in Fock space, and the Shannon entropy, measuring the degree of delocalisation of the eigenstates. The central question is whether scrambling the real-space structure while preserving the coarse organisation into neighbouring magnetisation sectors changes the phenomenology. As we show below, the two models behave qualitatively similarly under both diagnostics. Note that the two diagnostics are computed at different parameter values: the dynamical participation ratio is computed at $w=0$ (no Fock-space disorder) to isolate the effect of the kinetic constraint, while the Shannon entropy analysis uses $w=1$ to access the bulk of the spectrum where the transition of the East model is known to occur~\cite{Roy2019a}.

\subsection{Dynamical participation ratios}
\label{sec:ipr}

\begin{figure}[t]
  \centering
  \includegraphics[width=1.\columnwidth]{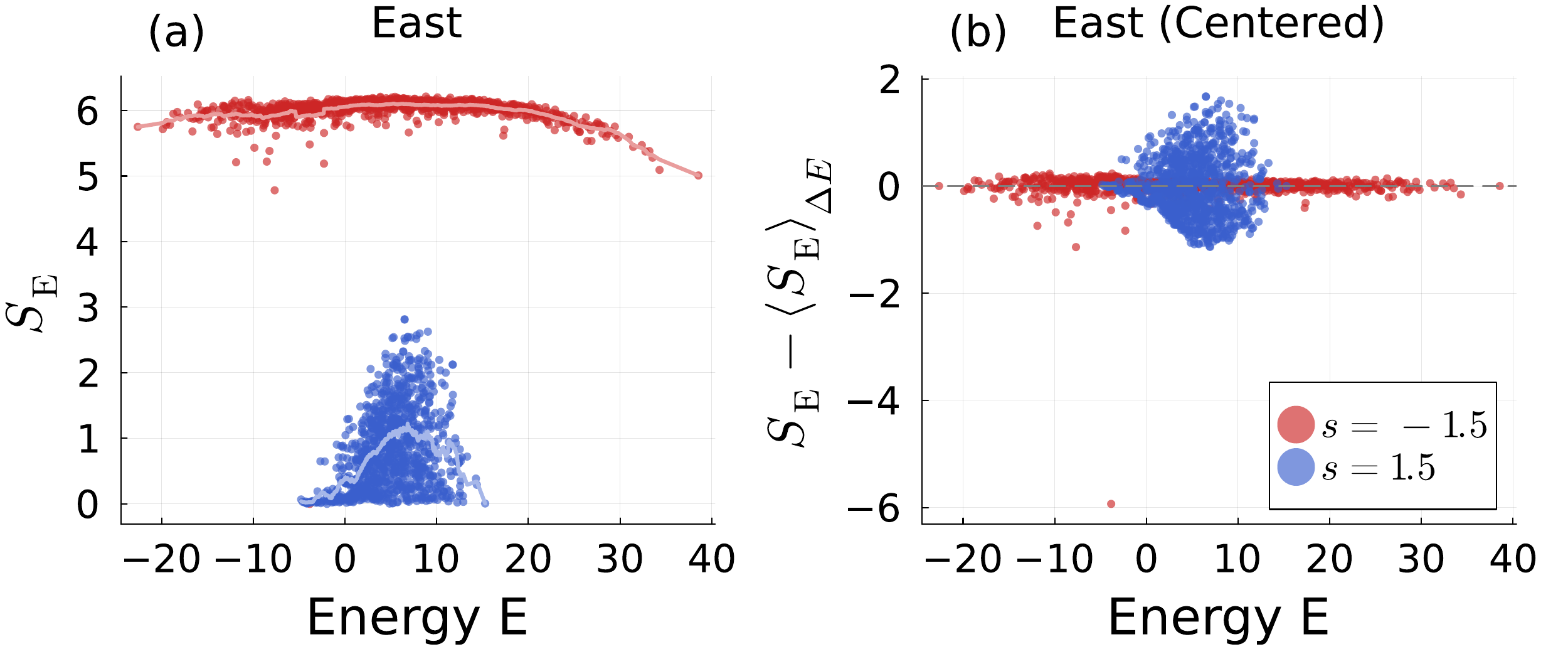}
  \caption{Representative examples of the Shannon entropy for each eigenstate of the East model for system size $L=10$ and $w=1$. Plots for the permuted model are similar, and data for both models is analysed in Fig.~\ref{fig:shannon-var}. (a) Entropy of each eigenstate versus its energy, for two values of $s$ each represented by a different colour. The lines in a lighter shade of each colour show the local averages over a small window in energy. Note how the fluctuations of the larger $s$ data around the mean are larger, as expected for a system violating ETH (see text). (b) Same data but with the local mean over the window $\Delta E$ subtracted to demonstrate fluctuations around the microcanonical average. A systematic analysis of this type of ETH violation is shown in Fig.~\ref{fig:shannon-var} (see Sec.~\ref{sec:entropy}).}  
  \label{fig:shannon-eth}
\end{figure}

\begin{figure}[t]
  \centering
  \includegraphics[width=1.\columnwidth]{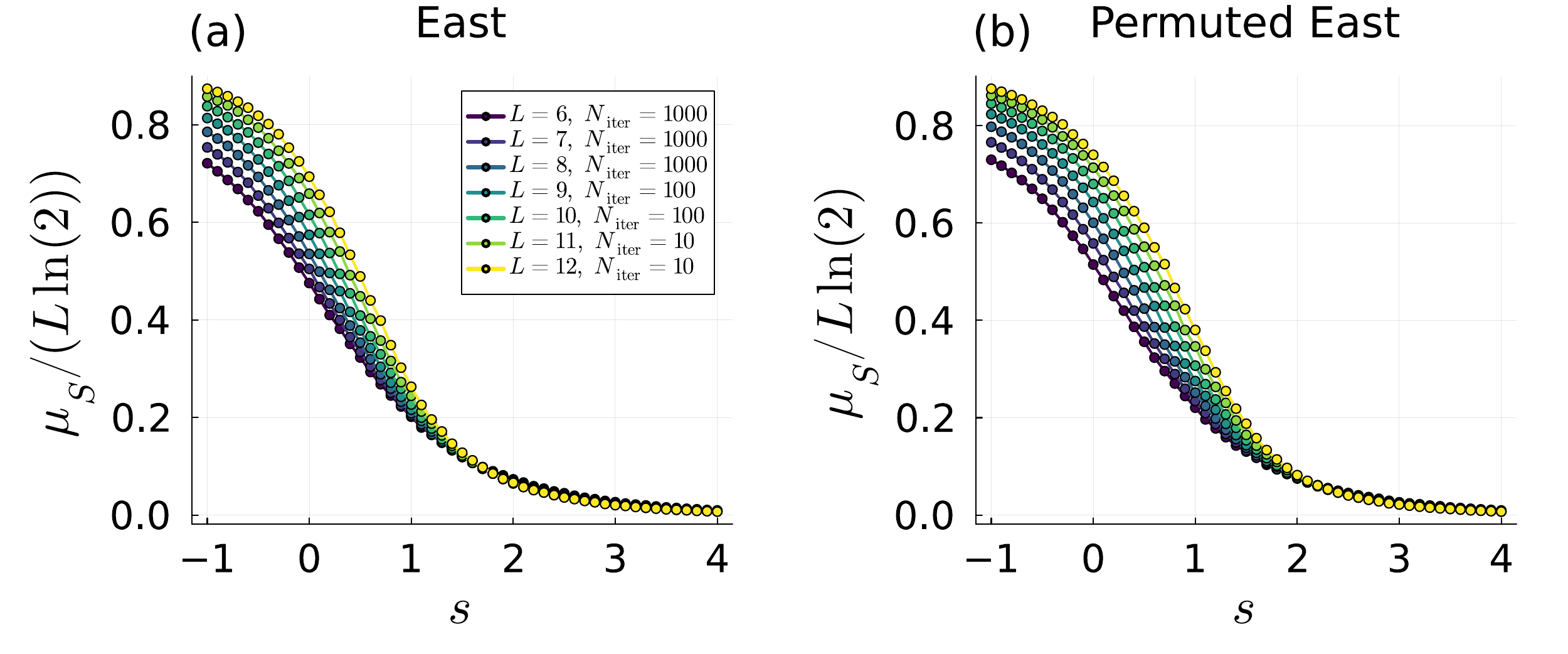}
  \caption{Mean scaled Shannon entropy $\mu_S/\ln D$ for the East (panel (a)) and permuted East (b) models, averaged over all eigenstates and disorder realisations (number of realisations indicated in the legend) for each size $L$. Here $w=1$. As $s$ increases the mean decreases, consistent with increasing localisation. For a quantitative determination of the transition point see Fig.~\ref{fig:shannon-var}.
  }
  \label{fig:shannon-mean}
\end{figure}

We probe Fock-space spreading via the dynamical participation ratio
\begin{equation}
   \phi(t) = \sum_m \left|\left< m\right.\left|\psi(t)\right> \right|^4
   \label{eq:phi-defn}
\end{equation}
where $\ket{m}$ are again Fock states. We initialise the system in each of the $L$ product states with exactly one spin up, and average the infinite-time limit, $\phi_\infty = \lim_{t\to\infty}\phi(t)$, over these $L$ initial states and, for the permuted East model, over realisations of the random permutation. In the extreme localised (in Fock space) case, where $\ket{\psi(t)}$ only has support on $\mathcal{O}(1)$ Fock states, $\phi(t)=\mathcal{O}(1)$, independent of system size or dimension of Hilbert space. On the other hand in the extreme delocalised case, $\phi(t)=\mathcal{O}(D^{-1})$ as the state has support on $\mathcal{O}(D)$ states (where $D=2^L$ is the Hilbert space dimension). Note that the latter extreme case differs from localisation in interacting systems due to disorder in real space, for which multifractality is expected~\cite{evers_anderson_2008,DeLuca2013,mace_multifractal_2019}. As here we are interested in a comparison between East and permuted East models and the unconstrained Ising model, we will not distinguish between the two cases.

The infinite-time average of this quantity, $\phi_\infty$, can be directly obtained in the absence of degeneracies in the standard way~\cite{Reimann:2008hq} as
\begin{equation}
   \phi_\infty
   = 2\sum_n \Bigl(\sum_\alpha |c_\alpha|^2\,|\langle n|\alpha\rangle|^2\Bigr)^{2}
   - \sum_{n,\alpha}|c_\alpha|^4\,|\langle n|\alpha\rangle|^4.
   \label{eq:phi-avg}
\end{equation}
where $c_\alpha = \bra{\alpha}\left.\psi(0)\right>$ is the overlap of the initial state with the $\alpha$-th eigenstate.

The results are shown in Fig.~\ref{fig:ipr-both}. For $s<0$ all three models trend towards small (delocalised) values with system size. For $s>0$ the two constrained models trend towards $\mathcal{O}(1)$ values consistent with localisation, while the Ising model remains at values orders of magnitude smaller, consistent with it staying delocalised. The precise location of any transition cannot be extracted from these data alone due to finite-size noise, but the qualitative difference between $s<0$ and $s>0$ is clear.

\subsection{Shannon entropy}
\label{sec:entropy}

To probe the same phenomenology from the eigenstate perspective, we now turn to the Shannon entropy, which characterises how broadly each eigenstate is spread over the Fock-state basis.

\begin{figure*}[t]
  \centering
  \includegraphics[width=1.\textwidth]{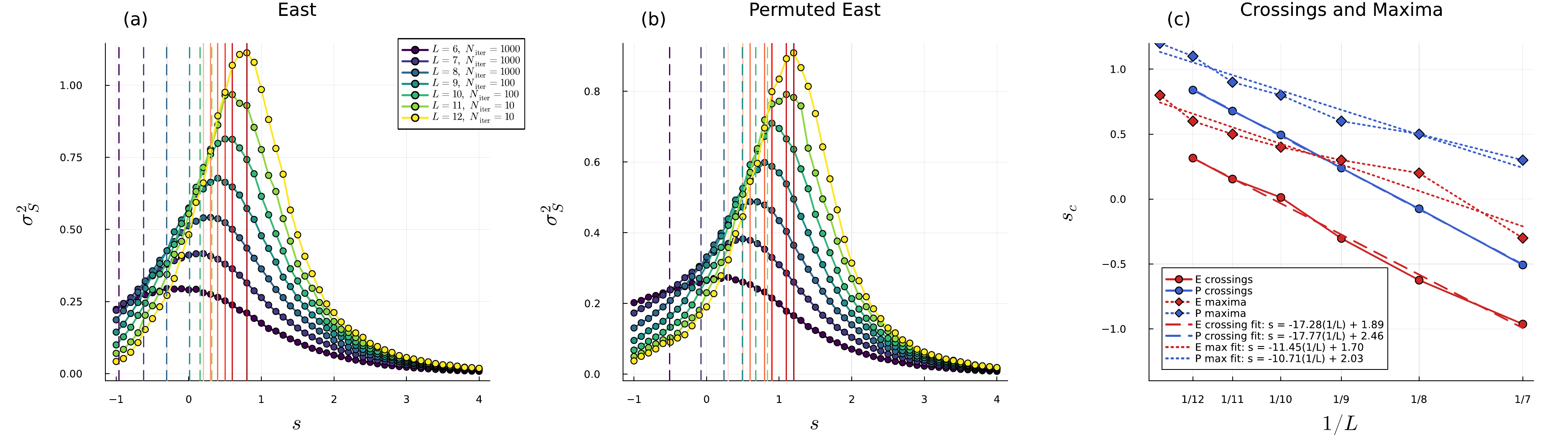}
  \caption{Local spectral variance $\sigma^2_S$ (Eq.~\ref{eq:shannon-var}), averaged over all eigenstates and disorder realisations (number of realisations indicated in the legend), for a number of system sizes $L$. Here $w=1$. Panels (a) and (b) show $\sigma^2_S$ for the East and permuted East models. Both display a reversal of the trend with system size at some $s$, indicating a transition from an delocalised to a localised phase, and a broad maximum in the variance, which provides a second finite-size estimate of the transition region. The crossing points and maxima both drift with $L$ and are therefore indicated for each pair of system sizes by dashed or solid lines, respectively. Panel (c) plots these (P and E refer to permuted and East models, respectively) versus $1/L$ along with a linear fit, showing that both extrapolate to a finite $s_c$ in the thermodynamic limit.}
  \label{fig:shannon-var}
\end{figure*}

For a state $\ket{\psi}$ and a complete basis $\ket{n}$, the Shannon entropy is 
\begin{equation}
   S_\psi=-\sum_n \left|c_n\right|^2\ln\left(\left|c_n\right|^2\right)
   \label{eq:shannon-entropy}
\end{equation}
where $c_n=\left<n\middle|  \psi\right>$. The entropy $S_\psi$ vanishes for a completely localised state and scales as $\ln(D)$ for one uniformly extended over all $D$ sites (in the present case $D=2^L$). The Shannon entropy is therefore a measure of localisation~\cite{Luitz:2015iv}. In what follows we use $\mu_S$ to denote the mean Shannon entropy, averaged over all eigenstates and disorder realisations, and $\sigma^2_S$ for the local spectral variance defined below.

As discussed in~\cite{DAlessio2016} and~\cite{Beugeling2014}, the eigenstate-to-eigenstate fluctuations of such measures decrease or increase with system size $L$ depending on whether the system is localised or not. This is analogous to the usual eigenstate thermalisation hypothesis (ETH) for local observables~\cite{Srednicki:1999bo}, and so we will also refer to the effect as violation of ETH. These fluctuations are also known to be largest near the transition~\cite{Bera2015}.

Fig.~\ref{fig:shannon-mean} shows the scaled mean entropy $\mu_S/\ln(D)$ for $w=1$, averaged over all eigenstates and over the number of realisations indicated in the legend. For both the East and permuted East models this quantity decreases as $s$ increases, consistent with increasing localisation in Fock space. In both cases the curves for different $L$ come close to crossing, which suggests a change in the finite-size flow, but the crossings are not sharp enough to locate the transition reliably from the mean alone.

To obtain clearer evidence for a transition, we turn to fluctuations around the microcanonical (MC) average. Fig.~\ref{fig:shannon-eth}(a) shows the Shannon entropy $S_E$ versus the energy for two representative values of $s$, with the MC local averages indicated by a continuous line, while Fig.~\ref{fig:shannon-eth}(b) shows the deviation from the MC average (that is, the difference between the entropy and the MC average). Here $\langle S_E\rangle_{\Delta E}$ denotes an average over eigenstates within a small energy window $\Delta E$ centred on the eigenstate in question. The broader distribution of the entropies for the larger value of $s$ is clearly visible in this plot.

To study this systematically, in Fig.~\ref{fig:shannon-var} we plot the local spectral variance
\begin{equation}
   \sigma^2_S \equiv \bigl\langle \bigl(S_E - \langle S_E\rangle_{\Delta E}\bigr)^2 \bigr\rangle_{\Delta E},
   \label{eq:shannon-var}
\end{equation}
where the average is taken over eigenstates within the energy window $\Delta E$ centred on the eigenstate in question, with $\Delta E$ set to 5\% of the total spectral bandwidth. This quantity is then averaged over all eigenstates and over the number of disorder realisations indicated in the legend, for a number of system sizes $L$.

The expected behaviour is that this variance vanishes in the thermodynamic limit for delocalised systems, but remains finite for localised systems~\cite{DAlessio2016,Beugeling2014}. The position of its maximum is also expected to approach that of the transition~\cite{Bera2015}. The variance therefore provides two finite-size estimators of the transition region. First, crossings between neighbouring system sizes identify the value of $s$ at which the finite-size flow changes sign: for small $s$ the fluctuations decrease with $L$, as expected in an delocalised phase, whereas for larger $s$ they remain finite, consistent with localised behaviour. Second, the maxima of the variance are susceptibility-like pseudocritical points~\cite{Igloi2007}\footnote{By pseudocriticality we mean the standard finite-size situation in which a feature such as a peak position defines an $L$-dependent estimate $s_c(L)$ that drifts with system size and approaches the true critical point only in the thermodynamic limit.}, identifying where the fluctuations are largest at finite size.

These two estimators need not coincide at accessible system sizes, since they probe different aspects of the transition and are subject to different finite-size corrections. We therefore view them as complementary sequences rather than as two independent definitions of $s_c$. Figure~\ref{fig:shannon-var}(c) shows the corresponding estimates versus $1/L$ for both models. For the East model, linear extrapolation of the crossings and maxima gives $s_c=1.9$ and $s_c=1.7$, respectively, while for the permuted East model the corresponding extrapolations give $s_c=2.5$ and $s_c=2.0$. Taken together, these results indicate the same qualitative finite-$s$ change from delocalised to localised behaviour in both models, while differing quantitatively in the location and finite-size drift of the estimated transition region.

\section{Conclusions and outlook}

We have studied two constrained models: the East-type model with local spin-flipping, and a permuted version in which the block structure in Fock space is preserved but the hopping terms are randomised so that the model is no longer local in real space.

Mapping the Hamiltonian to a graph in Fock space offers a unifying viewpoint. From this perspective, the key observation is that the slowing of the dynamics and the onset of localisation persist when the graph is similarly connected, even though the two real-space models are very different and the permuted model is not local. Both the East and permuted East models show a change in the dynamical participation ratio, indicating suppressed spreading in Fock space, and both show corresponding changes in the Shannon entropy, indicating a restructuring of the eigenstates. The latter diagnostic furthermore allows an extrapolation consistent with a finite transition point in both models as the parameter $s$ controlling the spin-flip amplitude is varied.

We note one structural feature that may be relevant for analytical approaches: for the Ising model, to go from a given Fock state to another at Hamming distance $d$ requires exactly $d$ hops along edges, so graph distance and Hamming distance coincide. For the East model this is no longer true---for example, going from $\ket{10001}$ to $\ket{11000}$ requires 5 hops---because the constraint removes edges from the graph. This complicates application of the forward-scattering approximation (FSA)~\cite{Roy2019a,Laumann:2014ju,Baldwin:2015ur}, which uses the number of shortest paths at a given Hamming distance as input.

Possible future directions include applying analytical methods such as the FSA or cavity methods, which might offer more insight into the importance of each structural feature of the model (such as number of paths between each site, typical path lengths between two sites differing by a given magnetisation, or distributions of connectivities), as well as extending this approach of identifying structural features of $H$ in some basis and randomising the Hamiltonian while preserving them to other constrained models.

A natural continuation is to systematically remove further structure from the Hamiltonian. For example, one could additionally randomise which magnetisation sectors are connected, allowing $|\Delta M|>1$, and ask whether the transition survives. If it does not, this would identify the ladder-like organisation into neighbouring sectors as the essential structural ingredient rather than real-space locality alone. Constructing a hierarchy of models that progressively removes structural features would map out which combinations are necessary and sufficient for the transition.

The dynamical participation ratio captures the long-time limit through $\phi_\infty$, but the time scale on which it is reached may differ between the two models even if the limiting value is the same. The rate at which $\phi(t)$ spreads in the delocalised phase, or saturates in the localised phase, could reveal quantitative differences not visible in the steady state.

Mapping out $s_c(w)$ would clarify the respective roles of kinetic constraints and Fock-space disorder. In particular, it is plausible that the transition persists at $w=0$: the hierarchical graph structure imposed by the kinetic constraint is present regardless of the diagonal disorder, and the latter is known not to cause localisation on its own~\cite{roy_fock-space_2020}, suggesting the constraint is the driving mechanism.

\bibliography{ms}

\end{document}